\documentclass[preprint,preprintnumbers,amsmath,amssymb,nofootinbib]{revtex4}
\usepackage{graphicx}
\usepackage{bm}

\newcommand\beq{ \begin{eqnarray} }
\newcommand\eeq{ \end{eqnarray} }

\newcommand{\average}[1]{\ensuremath{\langle#1\rangle} }
\usepackage{here}

\begin{document}


\title{Abelian spatial string tension \\in finite temperature SU(2) gauge theory}
\author{Takashige Sekiguchi}
\email{b11d6a04@s.kochi-u.ac.jp}
\affiliation{Graduate School of Intergrated Arts and Science Kochi University, Kochi 780-8520,Japan}
\author{Katsuya Ishiguro}
\email{ishiguro@kochi-u.ac.jp}
\affiliation{Integrated Information Center, Kochi University, Kochi 780-8520, Japan}

\date{\today}

\begin{abstract}
We investigate Abelian and monopole contributions to spatial string tension in the deconfined phase of finite temperature SU(2) gauge theory without imposing any gauge fixing conditions.  Lattice calculations of non-Abelian and Abelian spatial string tensions from the Wilson action at gauge coupling $\beta=2.74$ and lattice volume $24^3\times{N_t}$ $(N_{t}=\left\{ 24,8,6,4,2)\right\}$ show that these string tensions agree with each other within error bars at any adopted value of $N_t$, which implies Abelian dominance.  From measurements of non-Abelian, Abelian, and monopole forces that arise from the corresponding spatial string tension, furthermore, we find the tendency that the monopole contribution to the spatial string tension can be almost as large as the non-Abelian and Abelian ones.  The temperature dependence of the calculated non-Abelian and Abelian spatial string tensions allows us to conclude that the concept of dimensional reduction holds both for non-Abelian and Abelian sectors at temperatures higher than twice the critical temperature.
\end{abstract}


\maketitle

\section{Introduction}
\label{sec:1}
In QCD, which describes the strong interactions, color confinement is one of the open problems of great importance.  As a possible mechanism that explains color confinement , the dual superconductor picture of the QCD vacuum has been suggested.\cite{super1,super2}  While ordinary superconductors arise from condensation of electron Cooper pairs, magnetic monopoles condensate in dual superconductors.  The resultant dual Meissner effect acts to expel color electric flux from the QCD vacuum and thus, in the presence of a pair of particles of opposite color charge, to confine the intervening electric flux to flux tubes in such a way as to make the color electric potential between the two particles linear with respect to interparticle spacing.  This linear potential can induce color confinement.
In contrast to field theories with scalar fields such as the George-Glashow model\cite{gg1,gg2} and SUSY QCD,\cite{susy} It is difficult to directly identify magnetic monopoles in QCD.  For the purpose of circumventing this situation, et Hooft put forth an idea of deriving magnetic monopoles of integer charge by reducing SU(3) gauge theory to [U(1)]$^2$ via partial gauge fixing.\cite{thooft}  So far a large number of investigations of color confinement based on this idea have been performed via lattice simulations.  Among others, calculations that adopt maximally Abelian (MA) gauge\cite{ma1,ma2} in partial gauge fixing and treat monopoles defined on lattice {\'a} la DeGrand-Toussaint\cite{monop} strongly support the dual superconductor picture of the QCD vacuum.\cite{suzuki93,singh93,ejiri97,chernodub97,bali98,suzuki98,koma03-1,koma03-2} For example, squeezing of color electric flux due to the dual Meissner effect,\cite{singh93,bali98,koma03-1,koma03-2,bali96} Abelian dominance responsible for color confinement,\cite{suzuki93,ejiri97,kitahara95} monopole dominance responsible for color confinement,\cite{ejiri97,stack94,shiba94,shiba95,suzuki95,ejiri95} and effective monopole action\cite{shiba95,shiba95-2,kato98,fujimoto00} have been obtained under MA gauge fixing condition.  These results suggest the existence of the dual Ginzburg-Landau type of effective action in the infrared region of QCD.\cite{ezawa82,suzuki88,maedan89}

The same kind of calculations as the above-mentioned zero-temperature ones have been done at finite temperature; the thermal behavior of the Polyakov loop,\cite{suzuki97} Abelian and monopole dominance with respect to critical indices etc.,\cite{ejiri97} and the existence of finite temperature effective monopole action\cite{ishiguro00} have been therefrom observed.\cite{kitahara95,ejiri95,hioki91} This implies that even at finite temperature, Abelian parts and monopoles play a significant role in non-perturbative QCD.

One of the quantities that characterize non-perturbative nature of finite temperature QCD is spatial string tension $\sigma_{sp}$.  This tension arises from a space-space area law obeyed by the space-like Wilson loop in the magnetic sector of QCD.  According to earlier investigations in SU(2) gauge theory\cite{bali93-1,bali93-2} and SU(3) gauge theory,\cite{karsch95,boyd96} $\sigma_{sp}$ behaves as $\sqrt{\sigma_{sp}(T)}\sim g^2(T)T$, which turns out to be relevant to the string tension in three-dimensional gauge theory.  Also, full calculations for 2-flavor QCD\cite{maezawa07, bornyakov09} and (2+1)-flavor QCD\cite{cheng08} beyond the quenched approximation shows that $\sigma_{sp}$ agrees well with that predicted from dimensionally reduced QCD at temperature up to the vicinity of the deconfinement transition temperature $T_c$.

In this work, we focus on the spatial string tension in finite temperature SU(2) gauge theory and investigate its Abelian and monopole contributions.  Under MA gauge fixing condition, these contributions are known to lead to Abelian dominance and monopole dominance even in the high temperature deconfined phase.\cite{ejiri96}  Recently, on the other hand, description of color confinement in the dual superconductor picture has been reinforced by the fact that just like the case of MA gauge fixing, the dual Meissner effect, Abelian dominance, and monopole dominance occur under unitary gauge conditions such as F12 gauge and Polyakov gauge\cite{sekido07} and in the absence of gauge fixing.\cite{suzuki08,suzuki09} This suggests that Abelian monopoles are effective at explaining color confinement in a gauge-independent manner.\cite{suzuki14}  Thus, we here measure Abelian and monopole contributions to the spatial string tension without imposing any gauge fixing conditions and investigate the question of whether or not such contributions incorporate non-perturbative nature into the high temperature deconfined phase.

The present paper is organized as follows.  In Sec.\ 2, a theoretical background is described.  Section 3 is devoted to description of computational details.  In Sec. \ 4, our conclusions and future plans are given.

\section{Theoretical Background}
\label{sec:2}
We start with writing down the SU(2) Wilson action as

\beq
S&=&\beta \sum_{s,\mu>\nu}\left[1-\frac{1}{4}\mathrm{Tr}(U_{\mu\nu}(s)+U^{\dagger}_{\mu\nu}(s))\right], \\
U_{\mu\nu}(s)&=&U_\mu(s)U_\nu(s+\hat{\mu})U_\mu^\dagger(s+\hat{\nu})U_\nu^\dagger(s).
\eeq
Here $\beta = 4/g^2$, and $U_\mu(s)$ is the non-Abelian link variable, which is written in terms of the Pauli matrices $\sigma^a$ $(a=1,2,3)$ as

\begin{eqnarray}
U_{\mu}(s)=U_\mu^0(s)+i\sigma^a \cdot U_\mu^a(s).
\end{eqnarray}
The Abelian link variable, $u_\mu^a(s)$, of color direction $a$ can be derived from the non-Abelian link variable as

\begin{eqnarray}
u_\mu^a(s)&=&\cos\theta_\mu^a(s)+i\sigma^a\sin\theta_\mu^a(s), \\
\theta_\mu^a(s)&=&\arctan\left(\frac{U_\mu^a(s)}{U_\mu^0(s)}\right) \ \ \
\left(\theta_\mu^a(s) \in (-\pi,\pi]\right).
\end{eqnarray}
From the Abelian link variable, the Abelian field strength tensor is defined as

\begin{eqnarray}
\Theta_{\mu\nu}^a(s)
&=&\theta_\mu^a(s)+\theta_\nu^a(s+\hat{\mu})-\theta_\mu^a(s+\hat{\nu})-\theta_\nu^a(s)
\\
&=& \bar{\Theta}_{\mu\nu}^a(s)+2\pi n_{\mu\nu}^a(s)
\ \ \ \left(\bar{\Theta}_{\mu\nu}^a(s)\in (-\pi,\pi]\right) \label{thetabar}.
\end{eqnarray}
Here, $n_{\mu\nu}(s)$ denotes the number of the Dirac strings that penetrate the plaquette.  Then, the monopole current is defined as\cite{monop}

\begin{eqnarray}
k^a_\mu(s)=\frac{1}{2}\epsilon_{\mu\nu\rho\sigma}\partial_\nu n^a_{\rho\sigma}(s+\hat{\mu}),
\end{eqnarray}
which satisfies the conservation law $\partial_\mu^\prime k^a_\mu(s)=0$.  Here, $\partial_\mu$ ($\partial_\mu^\prime$) denotes the forward (backward) derivative.

In the presence of a pair of particles of opposite color charge, color confinement would cause the Wilson loop to obey an area law with increasing interparticle spacing and hence the color electric potential between the two particles to be proportional to interparticle spacing.  The string tension, defined as the coefficient affixed to the linear potential term, can thus be regarded as characteristic of color confinement.  The string tension decreases with increasing temperature and eventually becomes zero at $T_c$ where the system undergoes transition from the confined phase to the deconfined phase.  Interestingly, even above $T_c$, the spatial string tension remains nonzero.

The spatial string tension is defined as follows.  From the non-Abelian space-like Wilson loop
\begin{eqnarray}
&W_{NA}&(R,R')=\frac{1}{2}\mathrm{Tr}\prod_{s,\mu \in C} U_{\mu}(s),
\end{eqnarray}
we obtain the pseudo potential as

\begin{eqnarray}
V_{sp}^{NA}(R)= \lim_{R'\to \infty}\ln \frac{\average{W_{NA}(R,R')}}{\average{W_{NA}(R,R'+1)}}\label{wv},
\end{eqnarray}
where $R$ and $R'$ denote the length of the two sides of the Wilson loop in different spatial directions.  Finally we calculate the spatial string tension by assuming the pseudo-potential to have the form of

\begin{eqnarray}
V_{sp}^{NA}(R)=\sigma_{sp}^{NA}R-\frac{C}{R}+V_0,
\label{fit}
\end{eqnarray}
where $\sigma_{sp}^{NA}$ is the spatial string tension, $C$ is the Coulomb coefficient, and $V_0$ is constant, and then comparing it with the computational results.

We proceed to consider the contribution of the Abelian link variable to the spatial string tension and then divide it into monopole and photon contributions.  Hereafter we will omit color indices because in the absence of gauge fixing there is no preferred direction in color space.  The Abelian space-like Wilson loop is defined from the Abelian link variable as

\begin{eqnarray}
W_A=\exp{\left\{i\sum_s J_\mu(s)\theta_\mu(s)\right\}},
\end{eqnarray}
where $J_\mu(s)$ is the external current that takes on a value of $\pm1$ along the space-like Wilson loop.  Since $J_\mu(s)$ is conserved, we can write it as $J_{\nu}(s)=\partial'_{\mu}M_{\mu\nu}(s)$ by using the antisymmetric tensor $M_{\mu\nu}(s)$ that takes on a value of $\pm1$ on the plane surrounded by the space-like Wilson loop.  Thus, we obtain

\begin{eqnarray}
W_A=\exp\left\{-\frac{i}{2}\sum_s\Theta_{\mu\nu}M_{\mu\nu}(s)\right\}.
\end{eqnarray}
Equation (\ref{thetabar}) allows us to divide the Abelian space-like Wilson loop as

\begin{eqnarray}
&W_A&=W_M\cdot{W_P},\\
&W_M&=\exp\left\{2\pi{i}\sum_{s,s'}k_{\beta}(s)D(s-s')\frac{1}{2}\epsilon_{\alpha\beta\rho\sigma}\partial_{\alpha}M_{\rho\sigma}(s') \right\},\\
&W_P&=\exp\left\{-i\sum_{s,s'}\partial'_{\mu}\bar{\Theta}_{\mu\nu}(s)D(s-s')j_{\nu}(s')\right\},
\end{eqnarray}
where $D(s)$ is the Coulomb propagator on the lattice, $W_M$ is the monopole space-like Wilson loop, and $W_P$ is the photon space-like Wilson loop.  We evaluate the pseudo potentials by measuring these space-like Wilson loops in Monte Carlo simulation as

\begin{eqnarray}
&V_{sp}^{A}&(R)= \lim_{R'\to \infty}\ln \frac{\average{W_{A}(R,R')}}{\average{W_{A}(R,R'+1)}}, \label{wa} \\
&V_{sp}^{M}&(R)= \lim_{R'\to \infty}\ln \frac{\average{W_{M}(R,R')}}{\average{W_{M}(R,R'+1)}}, \label{wm} \\
&V_{sp}^{P}&(R)= \lim_{R'\to \infty}\ln \frac{\average{W_{P}(R,R')}}{\average{W_{P}(R,R'+1)}}. \label{wp}
\end{eqnarray}
Finally, we obtain the Abelian, monopole, and photon contributions to the spatial string tension, $\sigma_{sp}^A$, $\sigma_{sp}^M$, and $\sigma_{sp}^P$, by parametrizing the evaluated pseudo potentials in the form of Eq.\ (\ref{fit}).

In this work, we focus on the role played by Abelian monopoles in finite temperature SU(2) gauge theory.  In particular, the question of how configurations of Abelian fields and monopoles contribute to the above-described spatial string tension in the  confined and deconfined phases is not obvious, although Abelian dominance and monopole dominance in the spatial string tension were observed in the case of MA gauge fixing.\cite{ejiri96}  Here we shall approach this question without imposing any partial gauge fixing, motivated by the fact that the dual superconductor picture holds\cite{sekido07,suzuki08,suzuki09} in the confined phase irrespective of whether the gauge is fixed or not.

\section{Numerical Calculations}
\label{sec:3}
We have performed Monte Carlo simulation of finite temperature SU(2) lattice gauge theory to obtain the spatial string tension in the confined and deconfined phases.  The action, gauge coupling, and lattice size adopted here is the Wilson action, $\beta=2.74$, and $24^3\times N_t$, where $N_t$ is chosen as $24,8,6,4,2$ in such a way as to allow us to examine the temperature dependence of the spatial string tension.  To decrease errors we have implemented link integration\cite{parisi83,bali95} with respect to the non-Abelian link variable, while to obtain the ground state we have utilized APE smearing.\cite{ape}  In calculating the Abelian Wilson loop with reasonable accuracy, furthermore, we have adopted a random gauge transformation.\cite{suzuki08}  In the course of smearing, for the non-Abelian link variable in a spatial direction, we have summed up clumps that extend in perpendicular directions including a temporal direction as\cite{bali93-3}
\begin{eqnarray}
U_\mu(s) \mapsto P_{SU(2)}[U_\mu(s)
+\alpha\sum_{\nu\neq \mu}\left\{U_\nu(s)U_\mu(s+\hat{\nu})U^\dagger_\nu(s+\hat{\mu})\right.  \nonumber \\
\left.+U^\dagger_{\nu}(s-\hat{\nu})U_\mu(s-\hat{\nu})U_\nu(s+\hat{\mu}-\hat{\nu})\right\}],
\end{eqnarray}
where $\alpha$ is the smearing parameter, and $P_{SU(2)}$ represents projection onto SU(2). The errors are estimated by the jackknife method. Details of the parameters used for the simulation are tabulated in TABLE \ref{ta1}.
\begin{table}[H]
\centering
\caption{Parameters used for the simulation.  $T/T_c$ denotes the temperature in units of $T_c$; $N_{\rm smea}$, the repetition number of smearing; $N_{\rm conf}$, the number of configurations; $N_{\rm rgt}$, the number of random gauge transformations.}
\begin{tabular}{cccccc}\hline
 lattice size & $T/T_c$ & $N_{\rm smea}$ & $\alpha$ & $N_{\rm conf}$ & $N_{\rm rgt}$\\\hline
${24^{3}\times24}$ & {0}& {50} & {0.4} & {6400} & {5000}\\
$24^{3}\times8$ & 2& 60 & 0.4 & 28000 & 5000\\
$24^{3}\times6$ & 2.67 & 40 & 0.3 & 35000 &  5000\\
$24^{3}\times4$ & 4 & 30 & 0.3 & 40998 & 5000\\
$24^{3}\times2$ & 8 & 10 & 0.3 & 148000 & 5000 \\\hline
\end{tabular} \label{ta1}
\end{table}

\subsection{Pseudo potential and spatial string tension}
The non-Abelian pseudo potential, Eq.\ (\ref{wv}), and the Abelian pseudo potential, Eq.\ (\ref{wa}), calculated here are exhibited in Figs. 1-3. Figure 1 shows the pseudo potentials as function of $R/a$ with the lattice spacing a at $T=0 (N_t=24)$.  The left and right sides of Fig.\ \ref{fg1} illustrate the pseudo potentials as function of $R/a$ with the lattice spacing $a$ at $T=2T_c$ $(N_t=8)$ and $T=2.67T_c$ $(N_t=6)$, respectively, whereas the left and right sides of Fig.\ \ref{fg2} illustrate those at $T=4T_c$ $(N_t=4)$ and $T=8T_c$ $(N_t=2)$, respectively.  These figures show that in all the cases considered here the non-Abelian and Abelian pseudo potentials are nearly the same in the slope of the linear term.

\begin{figure}[H]
\centering
\includegraphics[width=0.48\textwidth]{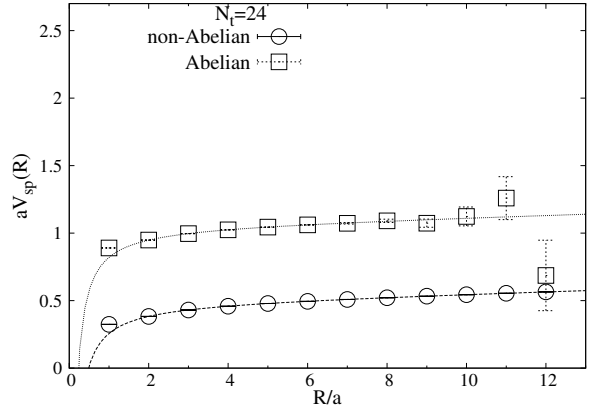}
\begin{flushleft}
\caption{The non-Abelian and Abelian pseudo potentials calculated as function of $R/a$ at $N_t=24$. The lines denote the best fitting by Eq.(\ref{fit}).}
\end{flushleft}
\label{fg0}
\end{figure}

\begin{figure}[H]
  \centering
\includegraphics[width=0.48\textwidth]{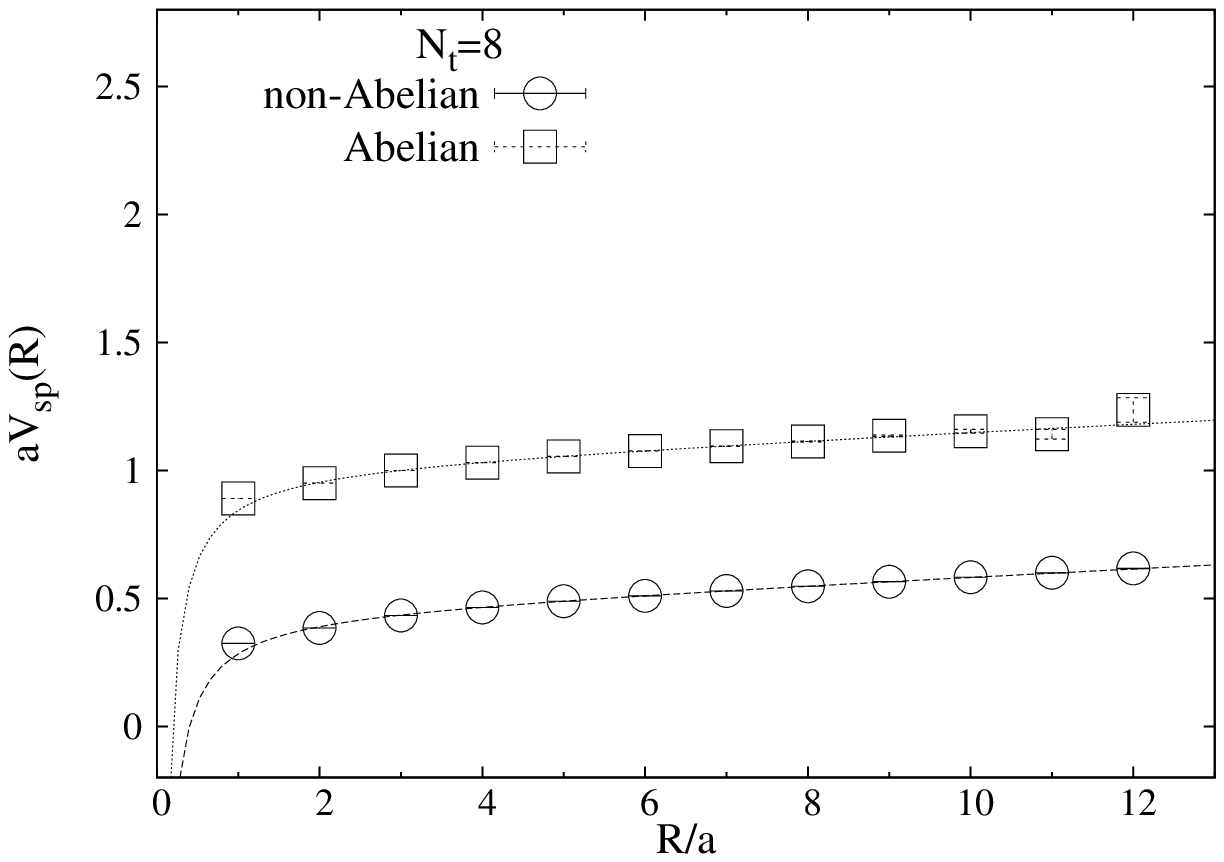}
\hspace{0.1cm}
\includegraphics[width=0.48\textwidth]{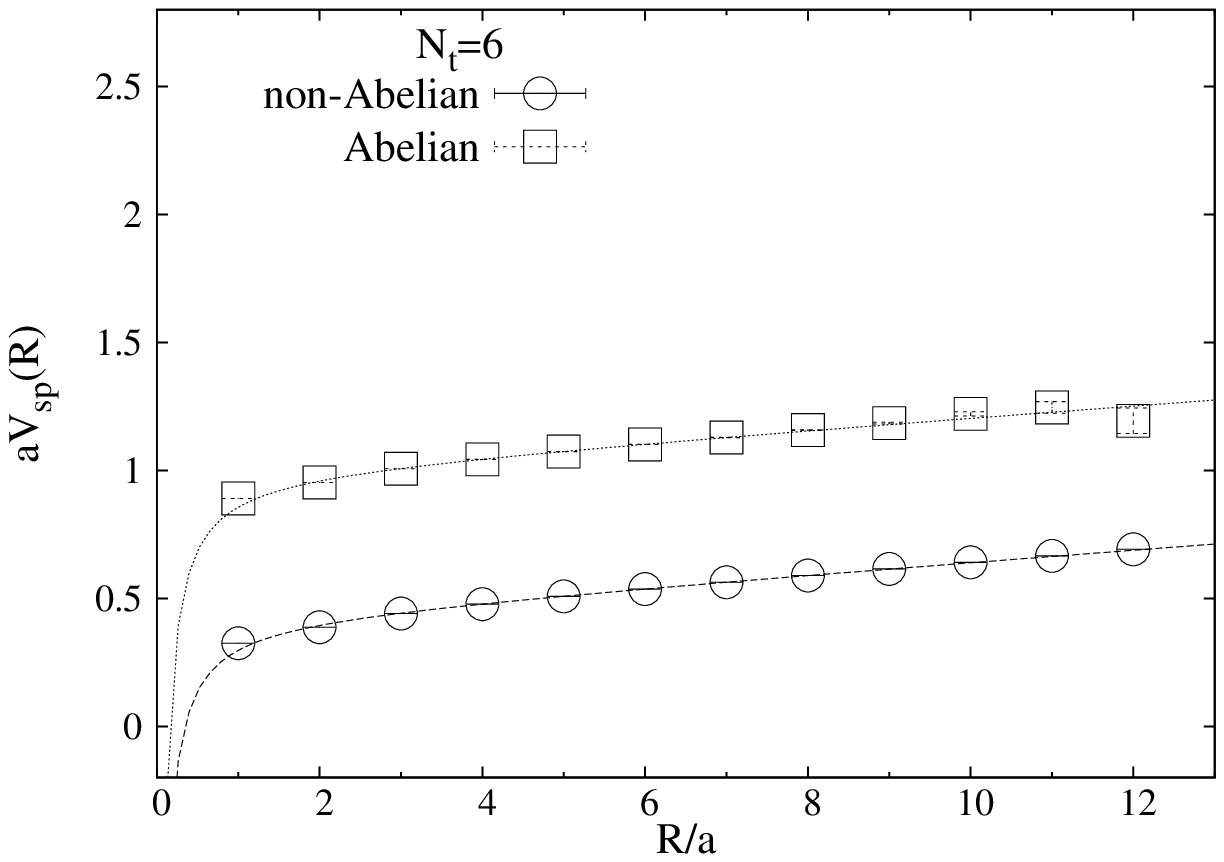}
\caption{Same as Fig. 1 at $N_t=8$ (left) and $N_t=6$ (right).   \label{fg1}}
\end{figure}

\begin{figure}[H]
  \centering
\includegraphics[width=0.48\textwidth]{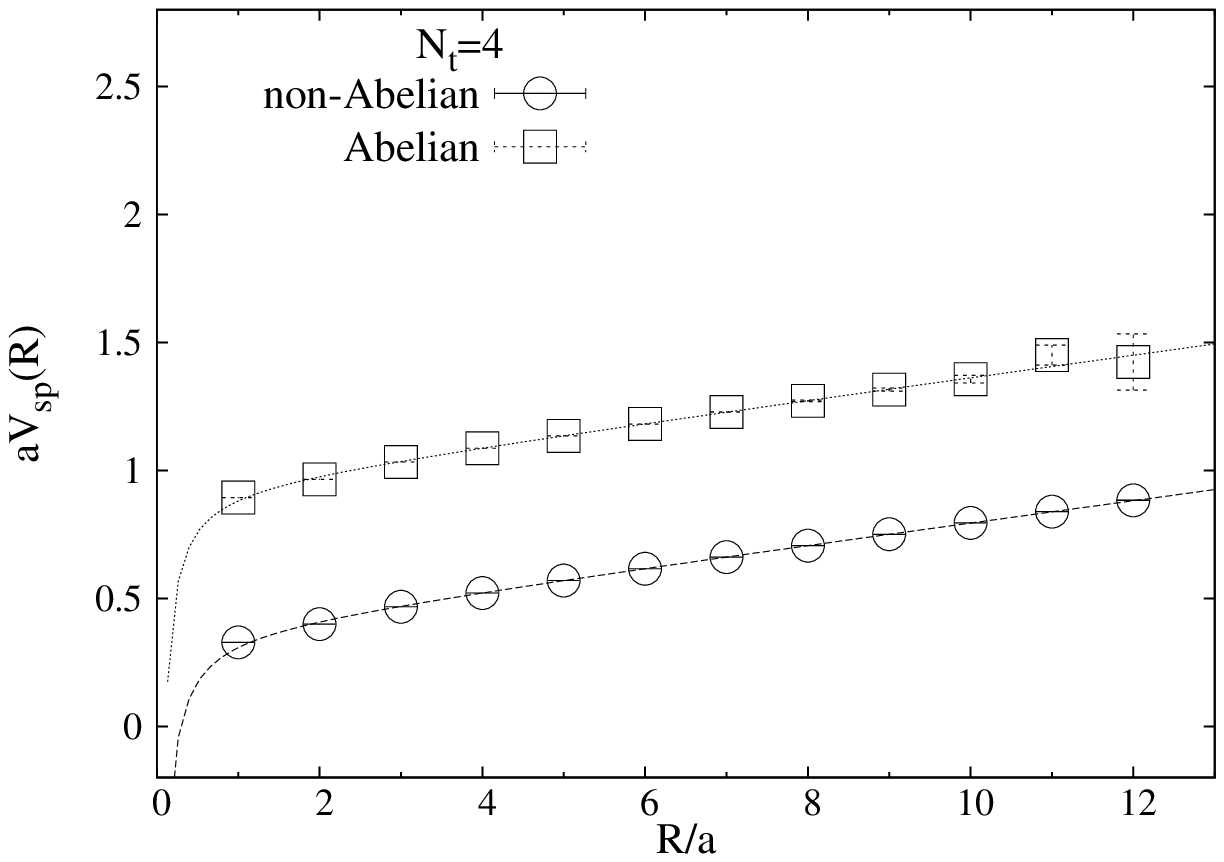}
\hspace{0.1cm}
\includegraphics[width=0.48\textwidth]{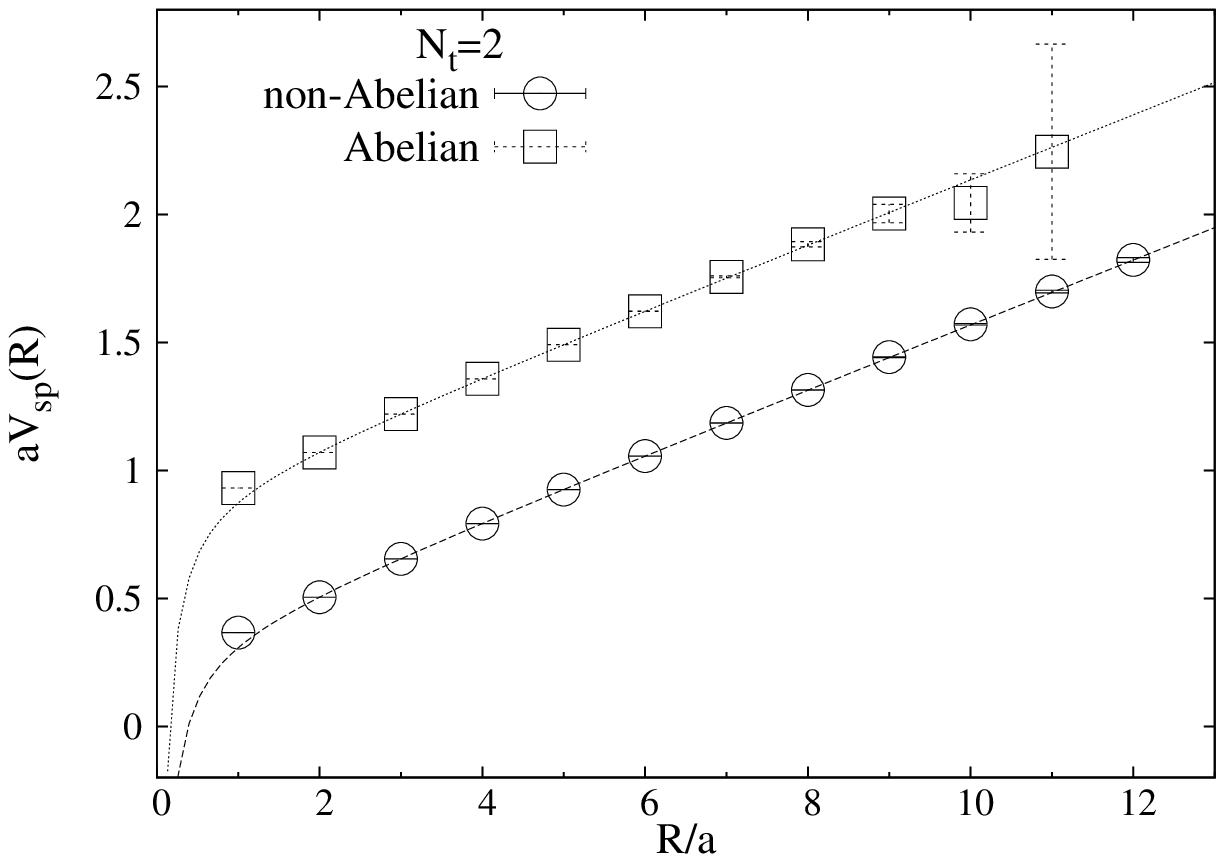}
\caption{Same as Fig. 1 at $N_t=4$ (left) and $N_t=2$ (right). \label{fg2}}
\end{figure}

The best fitting of the pseudo potentials exhibited in Figs.\ \ref{fg1} and \ref{fg2} leads to the non-Abelian spatial string tension shown in Table \ref{tas1} and the Abelian spatial string tension shown in Table \ref{tas2}.

\begin{table}[H]
\begin{minipage}{0.49\textwidth}
\caption{Fits to the non-Abelian pseudo potentials. Here, $R'/a$ corresponds to the optimal side length of the Wilson loop used for the smearing.}
\begin{tabular}{ccccc}\hline
 $N_t$ & $R'/a$ & $a^2\sigma^{NA}_{sp}$ & fit range &  $\chi^2/n.d.f$\\\hline
{24} & {5} & {0.0083(1)} & {4-8} & {0.0293}   \\
8 & 3 & 0.0150(1) & 4-8 & 1.0255   \\
6 & 2 & 0.0233(2) & 4-7  & 0.9056   \\
4 & 2 & 0.0427(1) & 4-9 & 1.0334  \\
2 & 3 &0.1257(1) & 2-8 & 0.9750  \\\hline
\end{tabular} \label{tas1}
\end{minipage}
\begin{minipage}{0.49\textwidth}
\caption{Same as Table \ref{tas1} for the Abelian pseudo potentials.}
\begin{tabular}{ccccc}\hline
 $N_t$ & $R'/a$ & $a^2\sigma^{NA}_{sp}$ & fit range &  $\chi^2/n.d.f$\\\hline
{24} & {4} & {0.0082(15)} & {4-8} & {0.1058}   \\
8 & 2 & 0.0147(4) & 4-9 & 0.8813   \\
6 & 2 & 0.0227(8) & 4-7  & 1.3468   \\
4 & 2 & 0.0433(7) & 4-9 & 0.8479  \\
2 & 2 &0.1260(2) & 2-8 & 0.9990  \\\hline
\end{tabular} \label{tas2}
\end{minipage}
\end{table}
As can be seen from Tables \ref{tas1} and \ref{tas2}, the Abelian spatial string tension agrees with the non-Abelian one within error bars.  In fact, as in Table \ref{si1}, $\sigma^A_{sp}/\sigma^{NA}_{sp}\approx 1$ in all the cases of $N_t$ considered here.  This result suggests that Abelian dominance holds for the spatial string tension even in the absence of gauge fixing. 

\begin{figure}[H]
\centerline{\includegraphics[width=80mm]{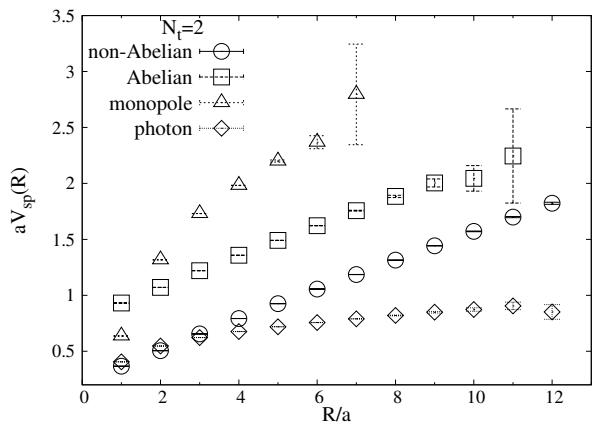}}
\caption{The non-Abelian, Abelian, monopole, and photon pseudo potentials at $N_t=2$ $(T=8T_c)$. \label{pf1}}
\end{figure}

\begin{table}[H]
\centering
\caption{The ratio of $\sigma^A_{sp}$ with $\sigma^{NA}_{sp}$.}
\begin{tabular}{ccc}\hline
 lattice size & $T/T_c$ & $\sigma^A_{sp}/\sigma^{NA}_{sp}$ \\\hline
 ${24^{3}\times24}$ & 0 & {0.99(18)}   \\
 $24^{3}\times8$ & 2     & 0.98(3) \\
 $24^{3}\times6$ & 2.67 & 0.97(3) \\
 $24^{3}\times4$ & 4     & 1.01(2) \\
 $24^{3}\times2$ & 8     & 1.00(2) \\\hline
\end{tabular}\label{si1}
\end{table}
We turn to the monopole and photon contributions to the spatial string tension. Figure \ref{pf1} exhibits the non-Abelian, Abelian, monopole, and photon pseudo potentials obtained from the corresponding space-like Wilson loops at $T=8T_c$ $(N_t=2)$.  As already seen from Fig.\ \ref{fg1} etc., we can determine the non-Abelian and Abelian spatial string tensions from fitting of the corresponding pseudo potentials with reasonable accuracy.  The monopole pseudo potential, however, contains too large errors at large $R/a$ for us to accurately determine the spatial string tension by fitting, but marginally shows a linear potential behavior.  We also find that the photon pseudo potential is consistent with zero spatial string tension and free from a linear potential behavior.
\subsection{Force}

Since, as mentioned above, the monopole contribution to the spatial string tension is hard to obtain from fitting of the monopole pseudo potential, we define the force $F_{sp}(R)$ from the incremental difference of the pseudo potential $V_{sp}(R)$ as

\begin{eqnarray}
F_{sp}(R)=\frac{1}{a}(V_{sp}(R)-V_{sp}(R-a)),
\end{eqnarray}
and, by substituting $V_{sp}^{NA}$, $V_{sp}^{A}$, and $V_{sp}^{M}$ into $V_{sp}$, measure the non-Abelian, Abelian, and monopole forces, respectively. The force, which reduces to the derivative of the pseudo potential in the continuum limit, keeps the spatial string tension, a coefficient affixed to a linear term of the pseudo potential, as a constant term; additional terms behave as $O(1/R^2)$ for large $R$.  The non-Abelian, Abelian, and monopole forces calculated for various lattice sizes are exhibited in Figs. 5-7.

\begin{figure}[H]
  \centering
%
\includegraphics[width=0.48\textwidth]{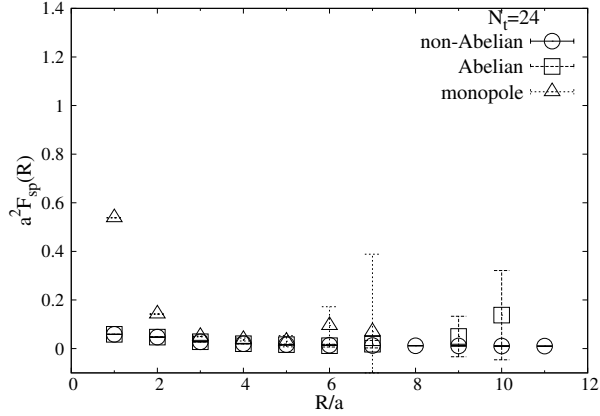}
\caption{The non-Abelian,Abelian,monopole forces at $N_t=24$.}
\label{fg00}
\end{figure}
\begin{figure}[H]
  \centering
\includegraphics[width=0.48\textwidth]{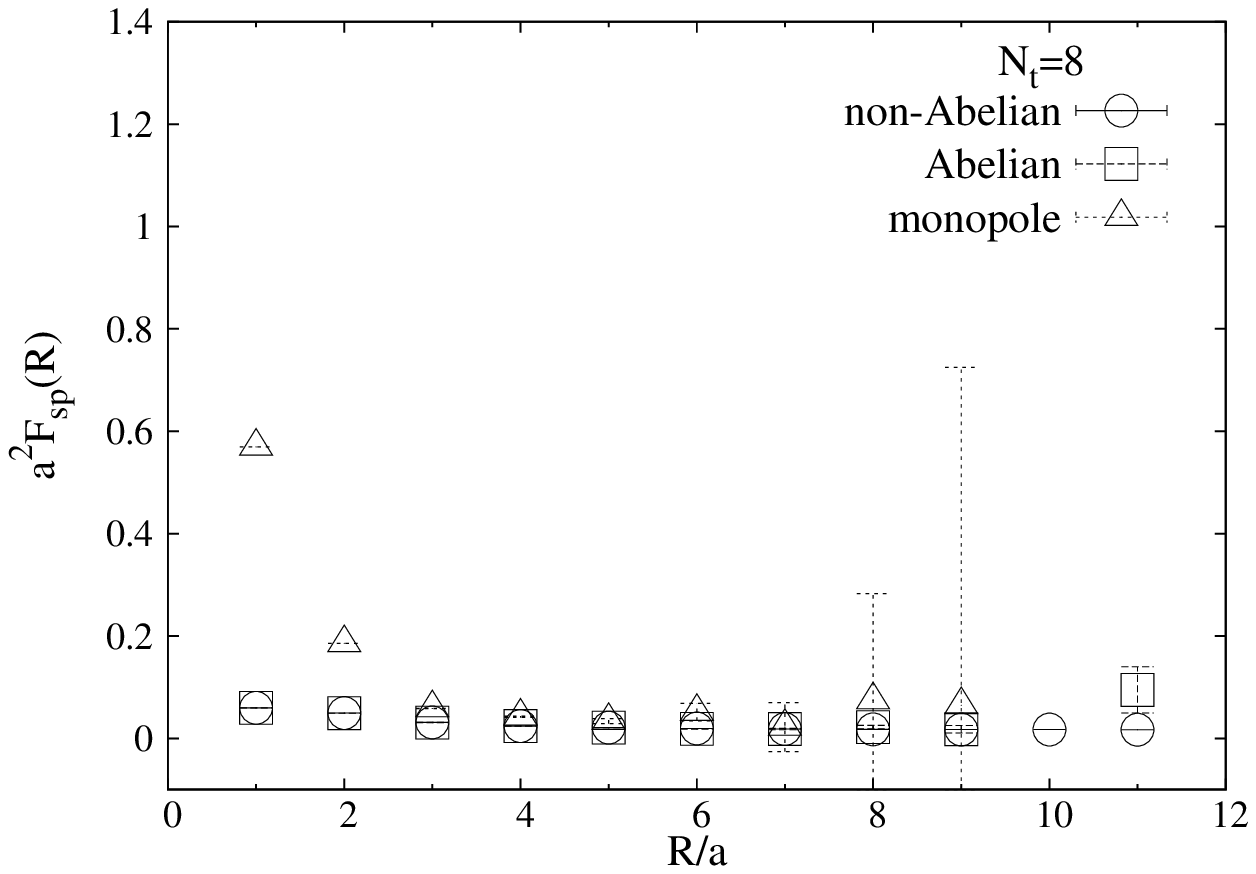}
\hspace{0.1cm}
\includegraphics[width=0.48\textwidth]{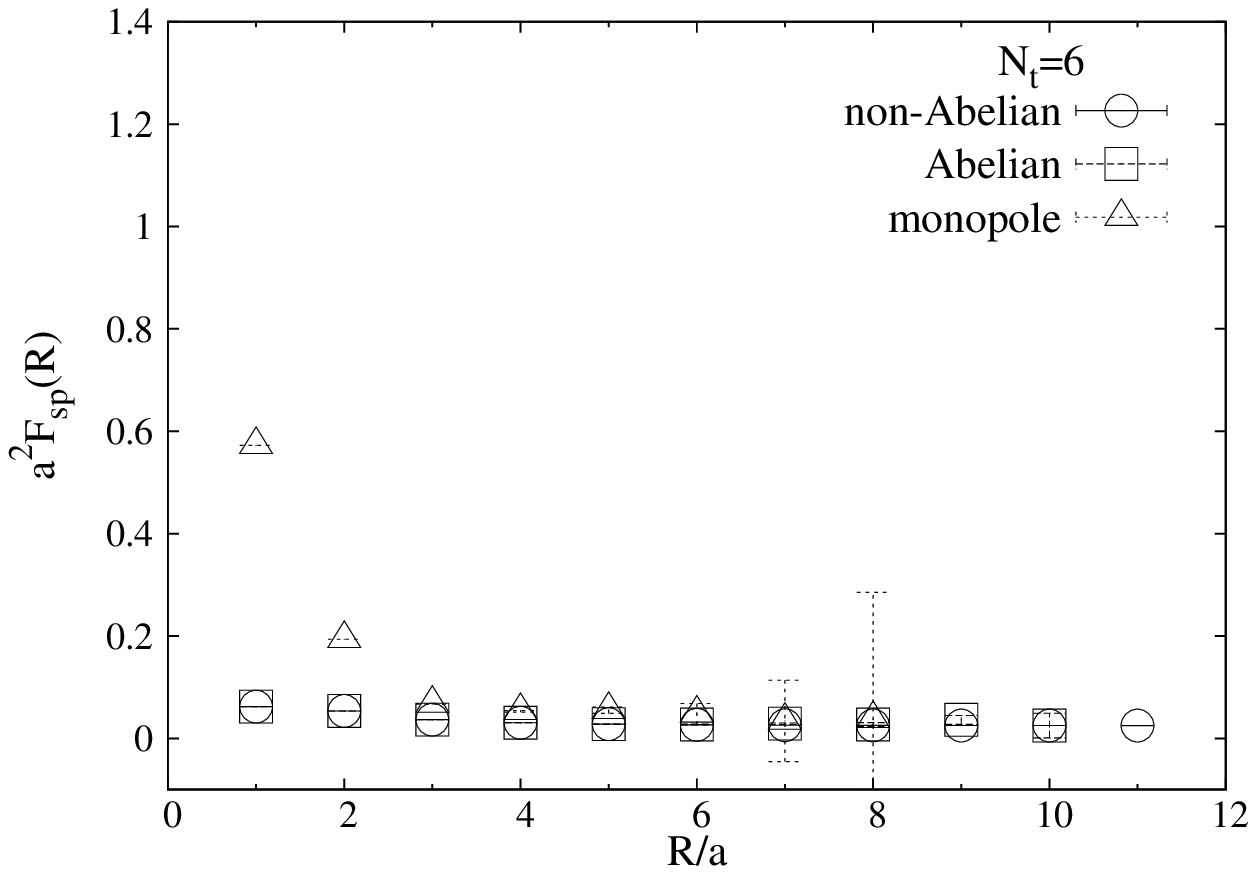}
\caption{ Same as Fig.5 at $N_t=8$ (left) and at $N_t=6$ (right). \label{ffa68}}
\end{figure}

\begin{figure}[H]
  \centering
\includegraphics[width=0.48\textwidth]{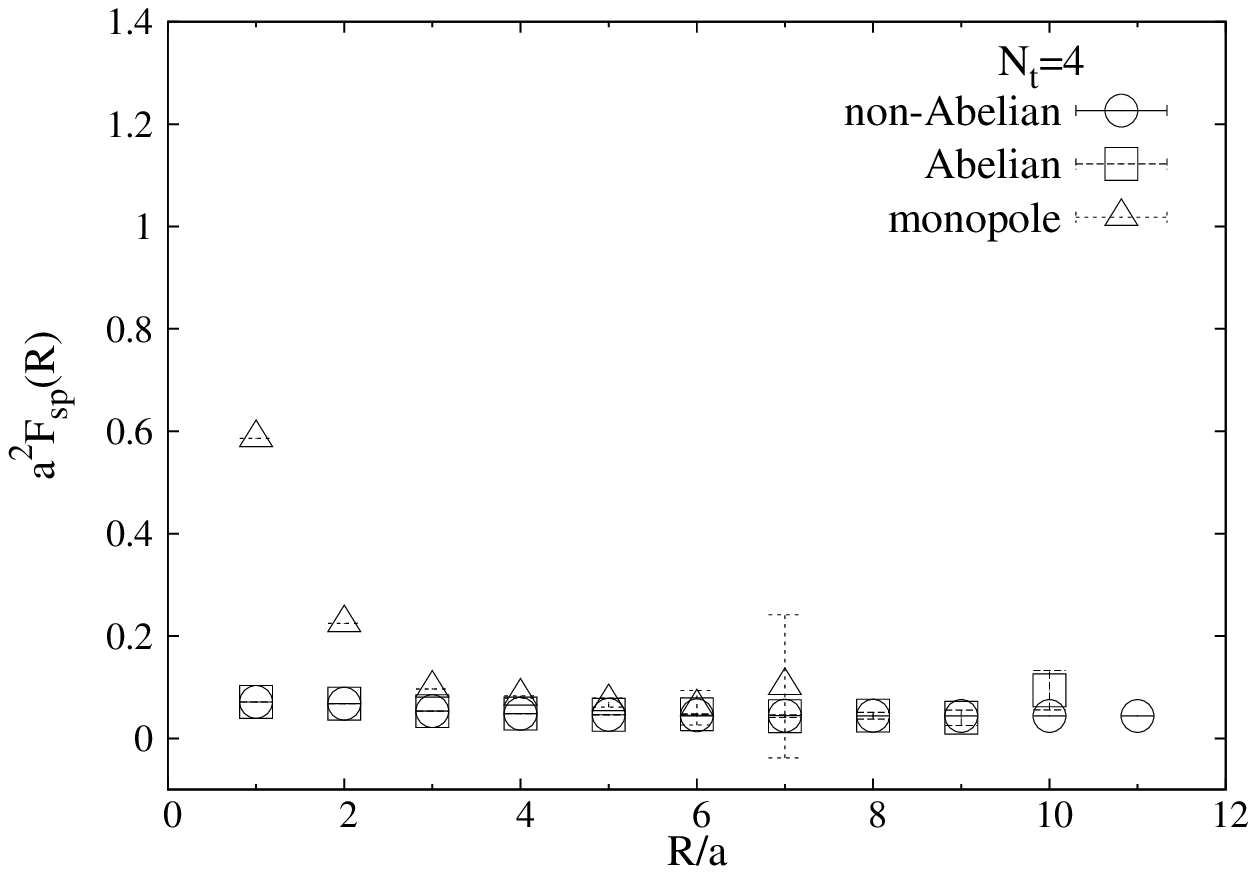}
\hspace{0.1cm}
\includegraphics[width=0.48\textwidth]{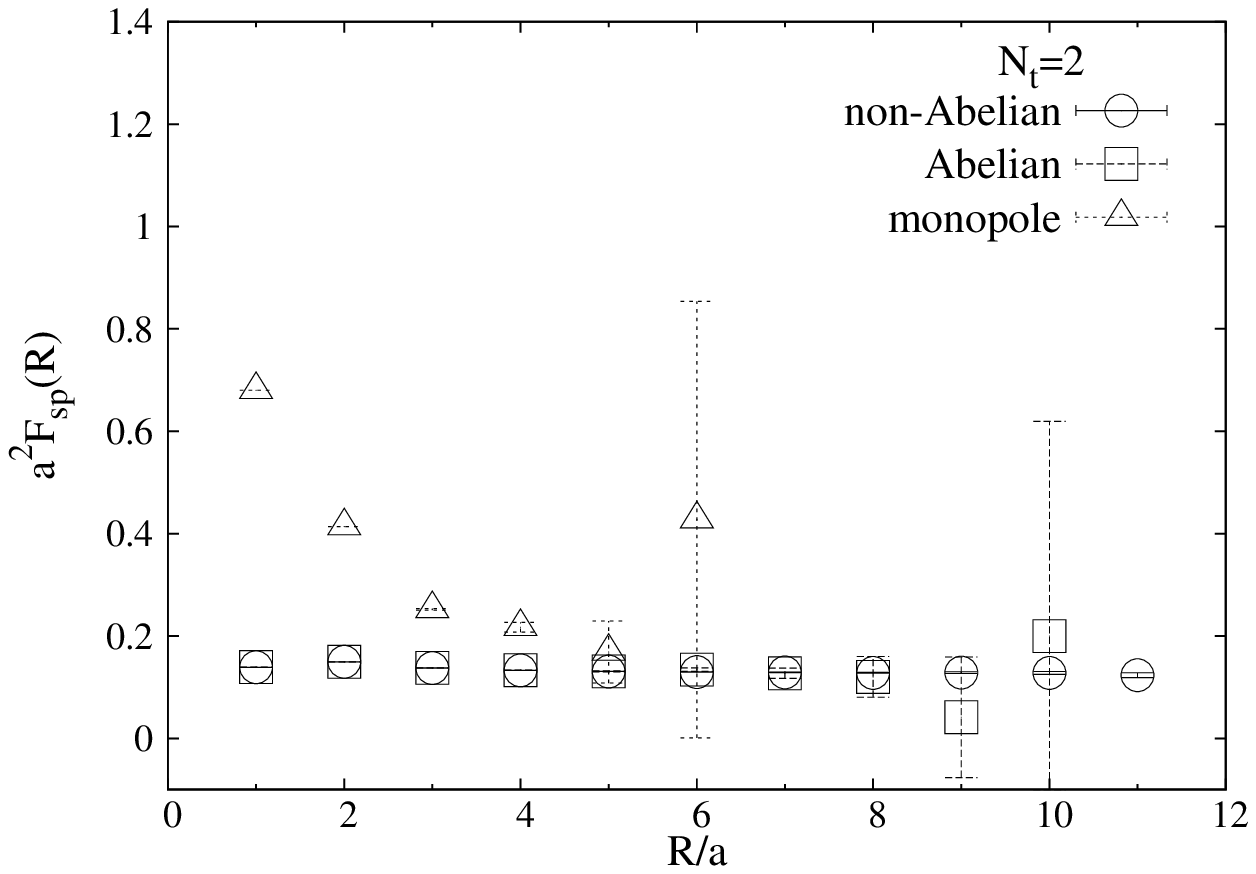}
\caption{Same as Fig.5 at $N_t=4$ (left) and at $N_t=2$ (right).  \label{ffa24}}
\end{figure}

We remark that while the calculated monopole pseudo potential contains too large errors at large $R$ for determination of the corresponding spatial string tension, the calculated non-Abelian, Abelian, and monopole forces at large $R$ agree with each other within error bars.  This result suggests that both in the confined and deconfined phases not only Abelian dominance but also monopole dominance holds for the spatial string tension even in the absence of gauge fixing.
\subsection{Temperature dependence}

\begin{figure}[H]
\centering
\includegraphics[width=80mm]{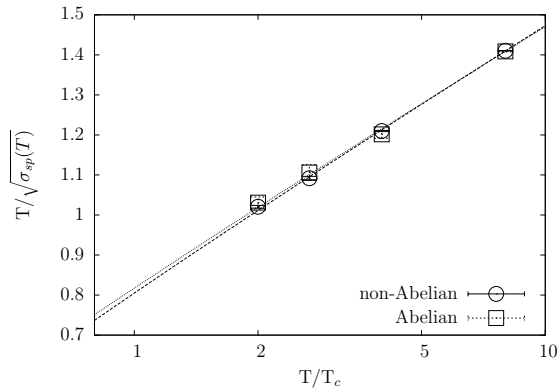}
\caption{The temperature dependence of the non-Abelian and Abelian spatial string tensions.  The lines denote the best fitting by Eq.\ (\ref{ggt}). \label{tdfa}}
\end{figure}

We now proceed to examine the temperature dependence of the spatial string tension.  The behavior of the spatial string tension at high temperature are predicted from the analogy of three-dimensional effective theory based on dimensional reduction as
\begin{eqnarray}
\sqrt{\sigma_{sp}(T)}=cg^2(T)T.
\label{ggt}
\end{eqnarray}
Here $c$ is constant, and $g(T)$ is the temperature-dependent four-dimensional running coupling constant that can be obtained in 2-loop perturbation theory as
\begin{eqnarray}
g^{-2}(T)=\frac{11}{12\pi^2}\ln\left(\frac{T}{\Lambda_T}\right)+\frac{17}{44\pi^2}\ln\left\{2\ln\left(\frac{T}{\Lambda_T}\right)\right\}.
\end{eqnarray}
The temperature dependence of the spatial string tension, which was obtained from fitting of the pseudo potentials as in Tables \ref{tas1} and \ref{tas2}, can be seen from Fig.\ \ref{tdfa} in which data are plotted on the $T/T_c$ vs.\ $T/\sqrt{\sigma_{sp}(T)}$ plane. 
The results for the fitting of the spatial string tensions by Eq.\ (\ref{ggt}) are tabulated in Table \ref{tasp}.
\begin{table}[H]
\centering
\caption{The parameters obtained from the fitting of the spatial string tensions by Eq.\ (\ref{ggt}).}
\begin{tabular}{ccc}\hline
 & $c$ & $\Lambda_{T}/T_c$ \\\hline
 non-Abelian& 0.360(5) & 0.087(5)  \\
 Abelian & 0.366(11) &  0.079(12) \\\hline
\end{tabular}\label{tasp}
\end{table}
We can observe from Table \ref{tasp} that the fitting parameters $c$ and $\Lambda_T$ are the same within error bars between the non-Abelian and Abelian cases.  That is, the Abelian spatial string has nearly the same temperature dependence as the non-Abelian one, and such dependence can be described by Eq.\ (\ref{ggt}) at $T \ge 2T_c$.
\subsection{Lattice spacing dependence }
So far we have fixed the gauge coupling $\beta$ at 2.74, which corresponds to fixed lattice spacing $a$. To confirm that the  
above-obtained results for the pseudo potentials and spatial string tensions do not depend on the lattice spacing, we apply reweighting\cite{swendsen88} to the data obtained for $\beta=2.74$.  We examine the non-Abelian and Abelian pseudo potentials and spatial string tensions by increasing $\beta$ from 2.75 to 2.76 by 0.01 for lattice size $24^3 \times N_t (N_t= 8,6,4,2)$.  Figure 9 illustrates the resultant spatial string tensions as function of $T/T_c$.  Here we have estimated $a$ for each $\beta$ from the non-perturbative $\beta$ function determined on lattice.\cite{engels95}
\begin{figure}[H]
\centering
%
\includegraphics[width=80mm]{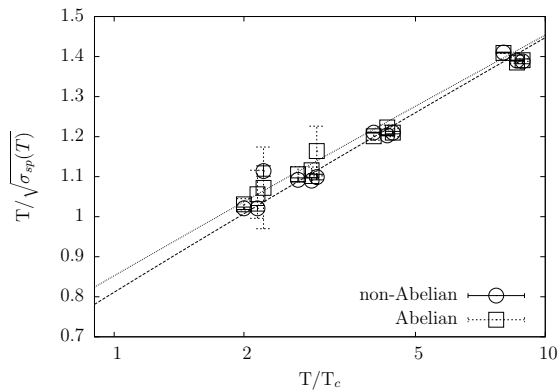}
\caption{Same as Fig.8, but the data obtained by reweighting are newly added.}
\label{rw}
\end{figure}
By including the data obtained by reweighting for $T\geq 2T_c$, we have updated the results for the fitting of the spatial string tensions by Eq. (22), which are tabulated in Table 6.  We find that both in the non-Abelian and Abelian cases, the adjustable parameters $c$ and $\Lambda_T$ in Table 6 agree with those in Table 5 within error bars.We can thus conclude that the spatial string tensions have no dependence on the lattice spacing. This suggests that the Abelian dominance holds for $T\geq 2T_c$ irrespective of lattice spacing.
\begin{table}[H]
\centering
\caption{Same as Table 5, but fitting is done by including the data obtained by reweighting.}
\begin{tabular}{ccc}\hline
 & $c$ & $\Lambda_{T}/T_c$ \\\hline
non-Abelian& 0.376(17) & 0.075(16)  \\
 Abelian & 0.394(44) &  0.056(19) \\\hline
\end{tabular}\label{tasprw}
\end{table}
Estimates of the monopole force by reweighting are generally difficult to make because large errors associated with reweighting are added to the errors inherent in the monopole force.  In fact, we are unable to make reasonable estimates of the monopole force for $\beta\geq2.77$.  We nevertheless have estimated the monopole force for $\beta=2.76$ by  
reweighting.  The results are shown for $N_t=8,6,4,2$ in Figs. 10 and 11, together with the non-Abelian and Abelian forces likewise estimated.  Within error bars, which are particularly large for the monopole force, these forces apparently agree with each other at least for large $R$.  This may imply that the monopole dominance, as well as  
the Abelian dominance, holds for $T\geq 2T_c$ irrespective of lattice spacing.
We remark in passing that shortage of available data in the confined phase ($N_t=24$) prevents us from estimating the pseudo potentials by reweighting.  The Abelian dominance in the confined phase, however, was established even in the absence of gauge fixing.\cite{suzuki08}  In fact, the data\cite{suzuki08} for the string tension obtained from the Wilson action at  
$\beta=2.5$ are effectively complemented by the present analysis.
\begin{figure}[H]
  \centering
\includegraphics[width=0.48\textwidth]{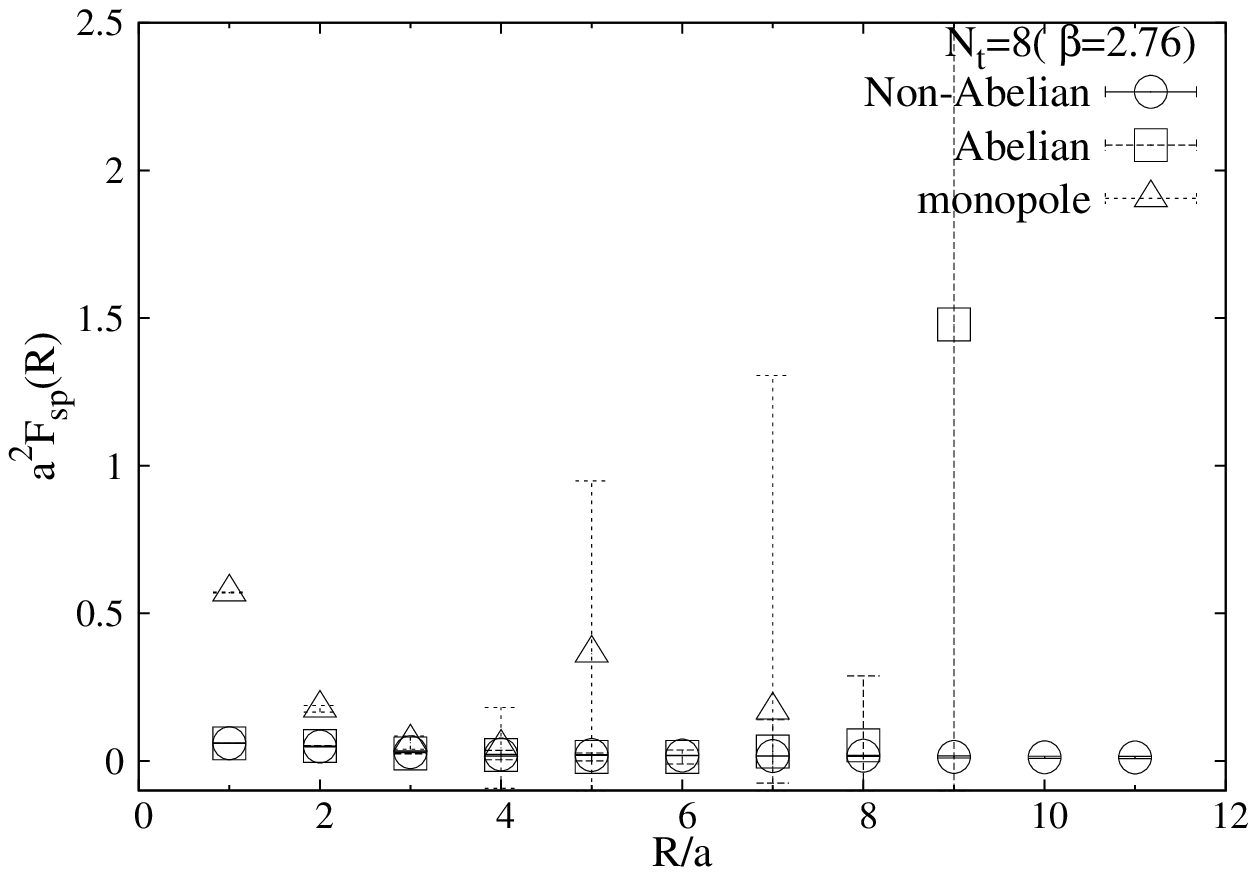}
\hspace{0.1cm}
\includegraphics[width=0.48\textwidth]{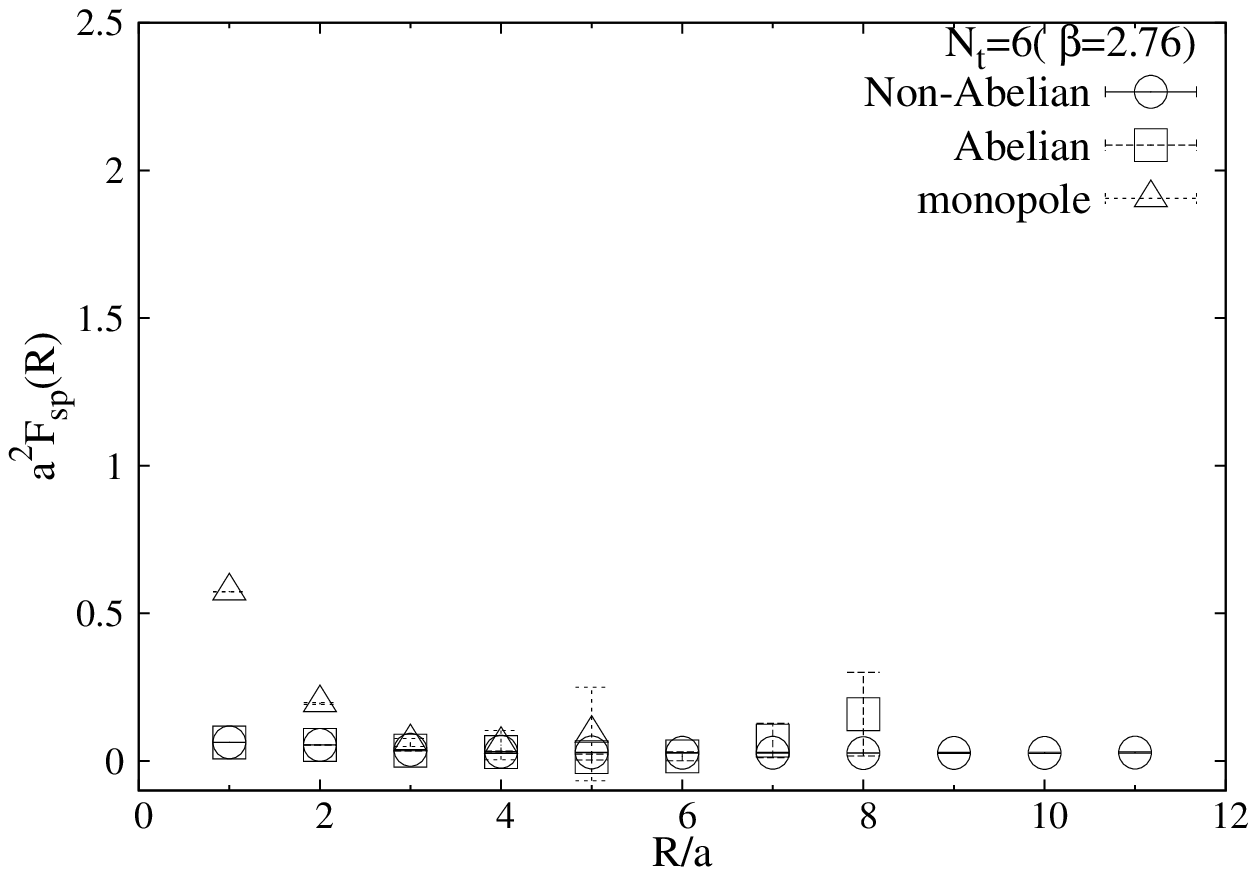}
\caption{The non-Abelian, Abelian, and monopole forces estimated from the data for $\beta=2.74$ by reweighting. The results for $\beta=2.76$ are plotted at $N_t=8$(left) and $N_t=6$(right).}
\label{t24b276f}
\end{figure}
\begin{figure}[H]
  \centering
\includegraphics[width=0.48\textwidth]{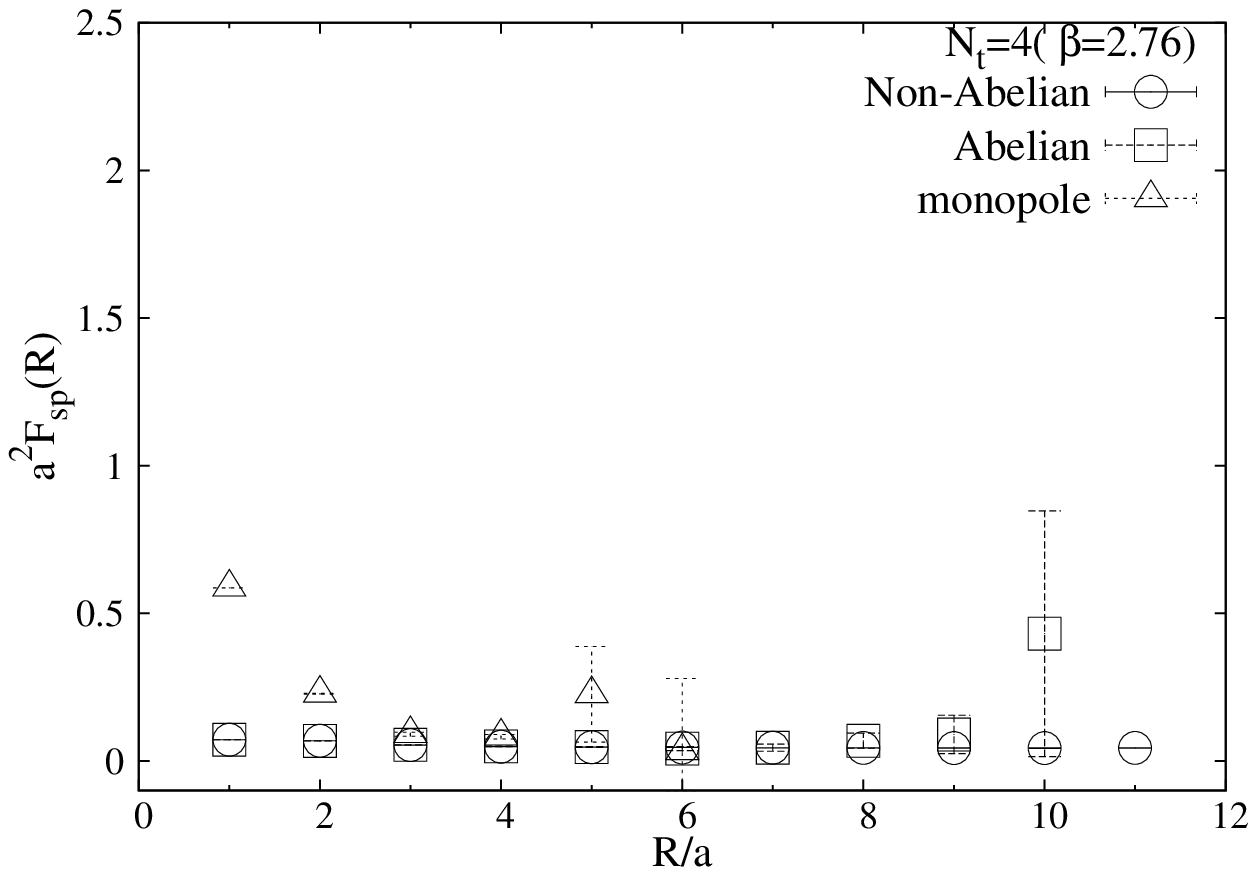}
\hspace{0.1cm}
\includegraphics[width=0.48\textwidth]{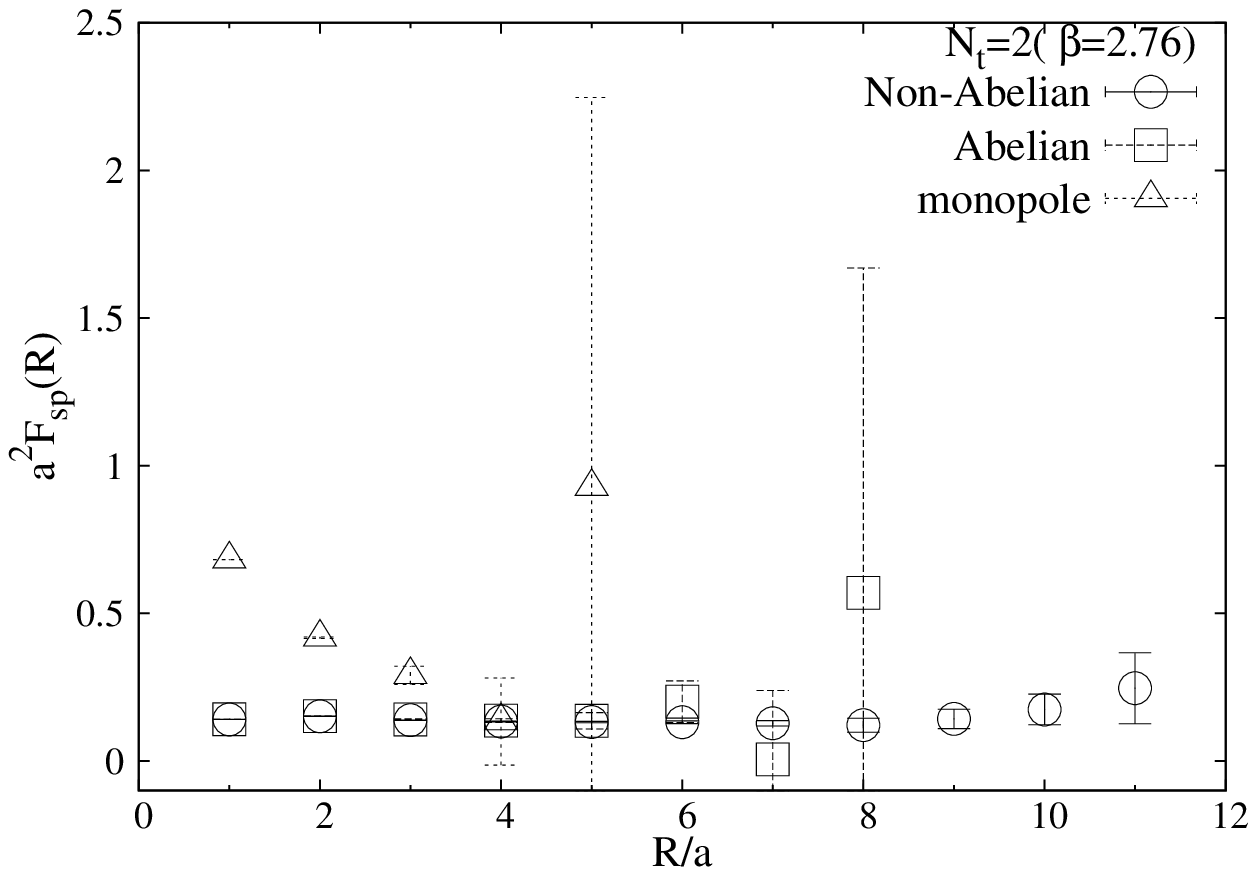}
\caption{Same as Fig.10 at $N_t=4$(left) and $N_t=2$(right).}
\label{t64b276f}
\end{figure}
\section{Concluding Remarks}
In this work we have focused on one of the non-perturbative quantities in finite-temperature SU(2) gauge theory, namely, the spatial string tension, and investigated the Abelian and monopole contributions without imposing any gauge fixing conditions.  The measurements have been performed by Monte Carlo simulation that adopts the Wilson action, $\beta=2.74$, and lattice size $24^3\times N_t$ $(N_t=\{24,8,6,4,2\})$.  The non-Abelian spatial string tension, which is gauge invariant, was investigated by Bali et al.,\cite{bali93-1} who confirmed the volume and lattice spacing independence of the spatial string tension for the parameters used in the present study. To examine the lattice spacing dependence, we apply reweighting to the data obtained for $\beta=2.74$. We have now found that for such parameters, the Abelian spatial string tension well reproduces the non-Abelian value, i.e., Abelian dominance holds irrespective of lattice spacing.  While the monopole contribution to the spatial string tension has yet to be evaluated from the fitting of the corresponding pseudo potential because the accuracy of the present calculations was not sufficient, we have discovered by measuring the forces between two particles of opposite color charge that at large interparticle spacing, the non-Abelian and monopole forces agree with each other within error bars.  This suggests that monopoles play an important role in the spatial string tension.  By examining the temperature dependence of the spatial string tension, furthermore, we have confirmed a good agreement with the behavior predicted from three-dimensional effective theory based on dimensional reduction at $T\ge 2T_c$ both in the non-Abelian and Abelian cases.  All these results used to be confirmed only under MA gauge fixing condition,\cite{ejiri96} and now have been reconfirmed in the absence of gauge fixing through new measurements of the Abelian and monopole contributions.
The monopole contribution to the spatial string tension remains to be determined with reasonable accuracy.  For such determination, higher statistics, extraction of wrapped monopoles,\cite{ejiri96} i.e., monopole loops closed by a temporal periodic boundary condition for finite-temperature systems, and so on would be helpful.  Since the present work suggests the possibility that magnetic monopoles play a non-perturbative role even in the deconfined phase in a manner that is independent of gauge fixing conditions, furthermore, investigations of how monopoles contribute to other non-perturbative quantities in the deconfined phase would give some insight to the still elusive physics of quark-gluon plasmas.
\section*{Acknowledgements}
This work was partially done by use of the High Performance Computing system and NEC SX-8R, SX-9, SX-ACE at Research Center for Nuclear Physics in Osaka University.
We would like to thank K. Iida for helpful discussions and his supports.





\end{document}